\documentclass[twocolumn,showpacs,pra,preprintnumbers]{revtex4}
\usepackage{amssymb}
\usepackage{amsmath}
\usepackage{graphicx}
\usepackage{dcolumn}
\usepackage{bm}
\usepackage{subfigure}
\usepackage{xcolor}
\usepackage[colorlinks,citecolor=blue, linkcolor=blue,hyperindex,dvipdfm]{hyperref}

\begin{document}

\title{Analytically computable symmetric quantum correlations}
\author{Li-qiang Zhang}
\author{Si-ren Yang}
\author{Chang-shui Yu}
\email{ycs@dlut.edu.cn}
\affiliation{School of Physics, Dalian University of Technology, Dalian 116024, China }

\begin{abstract}
One of the greatest challenges in developing the resource theory of a
quantum feature is to establish an analytically computable quantifier, which
directly limits the practicability of such quantifiers. Here, analytic
quantifiers of both the symmetric quantum discord (SQD) and the symmetric
measurement-induced nonlocality (SMIN) in a bipartite system of qubits are
studied on the basis of the quantum skew information. It is shown that the
SMIN of any two-qubit system and the SQD of bipartite "X"-type states and
block-diagonal states can be analytically determined. In addition, the SQD
and the SMIN are invariant with an attached quantum state. The validity of
our analytical expressions is further illustrated numerically on the basis
of several randomly generated density matrices.
\end{abstract}

\pacs{03.67.Mn, 03.65.Ud, 03.65.Ta}
\maketitle

\section{Introduction}

Quantum mechanical features are not only a fundamental aspect that
distinguishes the quantum from the classical world but also an important
physical resource for quantum information processing tasks (QIPTs). In
recent years, the resource theory of quantum features has been attracting
increasing interest (\cite{H1,H2,H3} and references therein); however, only
limited progress has been made toward the most essential goal, i.e., the
development of a good quantifier for a general quantum system. Quantum
entanglement is the most remarkable example; for this quantum feature, the
analytical quantifiers for a general quantum state are still restricted to
those for bipartite pure states \cite{B1,B2} and mixed states in six
dimensions (concurrence for qubit-qubit states\cite{e1} and negativity for
qubit-qutrit states \cite{N1,Horo,N2}), which were well established
approximately two decades ago. Needless to say, multipartite quantum states
include many inequivalent classes of entanglement \cite{ee3,miy}, and the
quantification of a bipartite high-dimensional state is usually related to
some complex optimization; consequently, it is difficult to formulate an
analytical or even an economic and effective numerical way to quantify the
entanglement of a general state. Therefore, the most fruitful approach is to
focus on effective quantification only for states in certain specific
classes \cite{CKW,book,adp,G3,b1,b2,G4,ycc,Ost,Y3,ee2,ee4}. In addition, as
a quantum correlation that extends beyond entanglement, the quantum discord
(QD) \cite{d1,d2}, which quantifies the discrepancy between the quantum
versions of two classically equivalent pieces of mutual information, can
only be analytically calculated for a general state in $2\otimes n$
dimensions \cite{NS,yzyz,L1,SY,RM,OUT,VON,QUAN,MULT,EXPE,QUANTU,X1,X2,X3}.
The same is true for its counterpart quantity \cite{loca,Wu,iajd}, that is,
the measurement-induced nonlocality (MIN) which characterizes the global
disturbance in a composite state caused by a local nondisturbing measurement
on one subsystem \cite{loca}. Recently, some progress has been made in
regard to analytical expressions for certain classes of states or effective
numerical methods \cite{iajd,L2,G1,KM,yb,S1,Hu,JOIN}. However, neither the
QD nor the MIN of a bipartite system is symmetric in the sense that
different results will be obtained (especially for some zero-discord states)
if the two subsystems are swapped, which implies that the QD and MIN cannot
completely quantify all of the quantum correlations present in a state. To
overcome this shortcoming, symmetric versions of the QD and the MIN are
defined, i.e., the symmetric quantum discord (SQD) \cite{Di1,Di2,Di3} and
the symmetric measurement-induced nonlocality (SMIN). In particular, with
the development of the resource theory of quantum coherence \cite%
{Y1,QQ1,c1,c2,c3,c4,c5,c6,c7,c8,c9,c10,QQ3,QQ2,M1,zhou,QUANTI,ACCE}, it was
shown that the coherence of a system could be converted through incoherent
operations into the SQD, instead of the asymmetric QD, of a composite system
\cite{QQ1,QQ2,QQ3}. By considering different bases, the coherence can also
be related to the SMIN. This finding suggests that these symmetric quantum
correlations could have fundamental meaning and potentially useful
applications. Outside of the context of the resource theory mentioned above,
much less progress has been made with regard to symmetric quantum
correlations, although it has been shown that the geometric SQDs of some
types of systems can be analytically calculated \cite{Di2}. However, the
geometric SQD seems not to be a good measure because forming a composite
state by taking the product with another mixed state will reduce the SQD of
the state of interest \cite{M1}.

In this paper, we study the analytical expressions for the SQD and SMIN of a
bipartite system of qubits. We define the SQD and SMIN in terms of the
quantum skew information \cite{skew,skew1,skew2,skew3} with regard to the
state of interest and projective measurements. It is obvious that neither
the SQD nor the SMIN will change if we take the product of another state
with the considered state based on the quantum skew information. Most
importantly, we find the analytical expression for the SMIN of any bipartite
state of qubits and find that the SQD can be analytically calculated for any
bipartite "X"-type states and block-diagonal states. As application
examples, we calculate both the SQD and the SMIN for several randomly
generated states. The numerical results are completely consistent with our
analytical expressions. The remainder of this paper is organized as follows.
In Sec. II, we introduce the definitions of the SQD and SMIN based on the
quantum skew information. In Sec. III, we present the analytical expressions
for both the SQD and SMIN as our main theorems. In Sec. IV, we consider
several randomly generated examples to test our analytical expressions. Sec.
V presents a discussion of our results and our conclusions.

\section{Definitions of the SQD and SMIN based on the skew information}

Before presenting the definitions of both the SQD and the SMIN, we would
like to introduce the definitions of the QD and MIN which will be given
based on the $l_{2}$ norm for the sake of intuitive understanding despite
the noncontractive nature of the $l_{2}$ norm. For a bipartite density
matrix $\rho _{AB}$, the (geometric) QD is defined as the minimal distance
away from the state that is disturbed by local projective measurements \cite%
{NS} as follows:%
\begin{equation}
\tilde{D}\left( \rho _{AB}\right) =\min_{\Pi ^{A}}\left\Vert \rho -\Pi
^{A}(\rho _{AB})\right\Vert _{2}^{2},  \label{QD}
\end{equation}%
where $\left\{ \Pi ^{A}\right\} $, defined as $\sum_{k}(\Pi _{k}^{A}\otimes
I^{B})\rho _{AB}(\Pi _{k}^{A}\otimes I^{B})$, denotes the projective
measurements on subsystem \textit{A} and $\left\Vert \cdot \right\Vert _{2}$
denotes the $l_{2}$ norm of a matrix. Similarly, the MIN is defined as the
maximal distance away from the state that is disturbed by local and \textit{%
subsystem-immune} projective measurements \cite{loca} as follows:
\begin{equation}
\tilde{N}(\rho _{AB})=\max_{\Pi ^{\prime A}}\left\Vert \rho -\Pi ^{\prime
A}(\rho _{AB})\right\Vert _{2}^{2},  \label{MIND}
\end{equation}%
where $\left\{ \Pi ^{\prime A}\right\} $ is defined similarly to $\left\{
\Pi ^{A}\right\} $ but requires $\sum_{k}\Pi _{k}^{\prime A}\rho _{A}\Pi
_{k}^{\prime A}=\rho _{A}$ to guarantee that the reduced density matrix $%
\rho _{A}=\mathrm{Tr}_{B}\rho _{AB}$ is not disturbed. It is apparent that
the QD and MIN are not symmetric since the projective measurements are
performed on only one subsystem. The two definitions above have a common
drawback, namely, they will changed if we take the product of an additional
mixed state with the state of interest; this change is directly induced by
the properties of the $l_{2}$ norm. This drawback would naturally be
inherited if we were to define the SQD and SMIN based on the $l_{2}$ norm.
Therefore, in what follows, we will present our definitions of the SQD and
SMIN in terms of the quantum skew information.

\textbf{Definition 1}.- For a bipartite $\left( m\otimes n\right) $%
-dimensional quantum state $\rho _{AB}$, the symmetric quantum discord (SQD)
based on the skew information is defined as
\begin{equation}
D(\rho _{AB})=\min_{\left\{ \left\vert k_{A}\right\rangle \right\} ,\left\{
\left\vert k_{B}\right\rangle \right\}
}\sum_{k_{A}=0}^{m-1}\sum_{k_{B}=0}^{n-1}I(\rho _{AB},K_{k_{A}k_{B}}),
\label{def1}
\end{equation}%
where%
\begin{eqnarray}
&&I(\rho _{AB},K_{k_{A}k_{B}})=-\frac{1}{2}\mathrm{Tr}[\sqrt{\rho _{AB}}%
,K_{k_{A}k_{B}}]^{2}  \notag \\
&=&\left\langle k_{A}k_{B}\right\vert \rho _{AB}\left\vert
k_{A}k_{B}\right\rangle -\left\vert \left\langle k_{A}k_{B}\right\vert \sqrt{%
\rho _{AB}}\left\vert k_{A}k_{B}\right\rangle \right\vert ^{2}  \label{skew}
\end{eqnarray}%
is the quantum skew information and $K_{k_{A}k_{B}}=\left\vert
k_{A}\right\rangle _{A}\left\langle k_{A}\right\vert \otimes $ $\left\vert
k_{B}\right\rangle _{B}\left\langle k_{B}\right\vert $ denotes the
projective measurements on both subsystems.

\textbf{Definition 2}.-The symmetric measurement-induced nonlocality (SMIN)
based on the skew information of the above quantum state $\rho _{AB}$, is
defined as
\begin{equation}
N(\rho _{AB})=\max_{\left\{ \left\vert k_{A}\right\rangle \right\} ,\left\{
\left\vert k_{B}\right\rangle \right\}
}\sum_{k_{A}=0}^{m-1}\sum_{k_{B}=0}^{n-1}I(\rho _{AB},K_{k_{A}k_{B}}),
\label{MINdingyi}
\end{equation}%
where $\left[ K_{k_{A}k_{B}},\rho _{A}\otimes \rho _{B}\right] =0$ with $%
\rho _{A\backslash B}=\mathrm{Tr}_{B\backslash A}\rho _{AB}$ denoting the
reduced density matrix.

It is obvious that both $D\left( \rho _{AB}\right) $ and $N\left( \rho
_{AB}\right) $ are symmetric if the two subsystems $A$ and $B$ are swapped.
In addition, both are invariant under local unitary transformations. One can
also see that both $D\left( \rho _{AB}\right) $ and $N\left( \rho
_{AB}\right) $ vanish if $\rho
_{AB}=\sum_{k_{A}}\sum_{k_{B}}P_{k_{A},k_{B}}\left\vert k_{A}\right\rangle
\left\langle k_{A}\right\vert \otimes \left\vert k_{B}\right\rangle
\left\langle k_{B}\right\vert $, where $\left\{ \left\vert
k_{A}\right\rangle \right\} $ and $\left\{ \left\vert k_{B}\right\rangle
\right\} $ are each some orthonormal (not necessarily complete) basis set.
These are the fundamental requirements for the symmetric versions of the QD
and MIN. Thus, based on the above definitions, we can present our main
theorems in the next section.

\section{Analytical expressions for the SQD and SMIN for two-Qubit systems}

\textbf{Theorem 1.-} Let $\rho _{A}$ and $\rho _{B}$ be the reduced density
matrices of the $2\otimes 2$-dimensional quantum state $\rho _{AB}$. Then,
the SMIN of $\rho _{AB}$ is directly given by
\begin{equation}
N(\rho _{AB})=\sum_{k_{A},k_{B}=0}^{1}I\left( \rho
_{AB},K_{k_{A},k_{B}}\right) ,  \label{nd}
\end{equation}%
if neither $\rho _{A}$ nor $\rho _{B}$ is degenerate (has the same nonzero
eigenvalues), or by
\begin{equation}
N(\rho _{AB})=1-\frac{1}{4}(\mathrm{Tr}\sqrt{\rho _{AB}})^{2}
\label{MINjiexi}
\end{equation}%
if both $\rho _{A}$ and $\rho _{B}$ are degenerate, or by%
\begin{equation}
N(\rho _{AB})=1-\frac{1}{2}\sum_{k=0}^{1}(\mathrm{Tr}\left( \left\langle
k\right\vert _{X}\sqrt{\rho _{AB}}\left\vert k\right\rangle _{X}\right) )^{2}
\label{M1}
\end{equation}%
if only one of $\rho _{A}$ and $\rho _{B}$ is degenerate with $\left\{
\left\vert k\right\rangle _{X}\right\} $ ($X=A$ or $B$), where $\left\{
\left\vert k\right\rangle _{X}\right\} $ denotes the orthonormal basis of
the nondegenerate subsystem.

\textbf{Proof}. If neither $\rho _{A}$ nor $\rho _{B}$ is degenerate,
namely, all eigenvalues of $\rho _{A}$ and $\rho _{B}$ are different from
each other, then their eigenvectors are uniquely determined. Thus, the only
projective measurement that does not disturb $\rho _{A}$ and $\rho _{B}$ is $%
K=\left\{ K_{k_{A}k_{B}}=\left\vert i\right\rangle \left\langle i\right\vert
\otimes \left\vert j\right\rangle \left\langle j\right\vert \right\} $,
where $\left\{ \left\vert i\right\rangle \right\} $ and $\left\{ \left\vert
j\right\rangle \right\} $ denote the eigenvectors of $\rho _{A}$ and $\rho
_{B}$, respectively. That is, the optimization in Equation (\ref{MINdingyi})
is not necessary, namely, Equation (\ref{nd}) holds.

If $\rho _{A}$ and $\rho _{B}$ are both degenerate, then the SMIN of $\rho
_{AB}$ in Equation (\ref{MINdingyi}) can be rewritten as
\begin{eqnarray}
N(\rho _{AB}) &=&\max_{\left\{ \left\vert k_{A}\right\rangle \right\}
,\left\{ \left\vert k_{B}\right\rangle \right\}
}\sum_{k_{A},k_{B}=0}^{1}\left\langle k_{A}\right\vert \left\langle
k_{B}\right\vert \rho _{AB}\left\vert k_{A}\right\rangle \left\vert
k_{B}\right\rangle   \notag \\
&&-\left\vert \left\langle k_{A}\right\vert \left\langle k_{B}\right\vert
\sqrt{\rho _{AB}}\left\vert k_{A}\right\rangle \left\vert k_{B}\right\rangle
\right\vert ^{2}  \notag \\
&=&1-\min_{\left\{ \left\vert k_{A}\right\rangle \right\} ,\left\{
\left\vert k_{B}\right\rangle \right\} }\sum_{k_{A},k_{B}=0}^{1}\left\vert
\left\langle k_{A}\right\vert \left\langle k_{B}\right\vert \sqrt{\rho _{AB}}%
\left\vert k_{A}\right\rangle \left\vert k_{B}\right\rangle \right\vert ^{2}
\notag \\
&=&1-\min_{U}\sum_{k=1}^{4}\left\vert U_{A}\otimes U_{B}\sqrt{\rho _{AB}}%
U_{A}^{\dagger }\otimes U_{B}^{\dagger }\right\vert _{kk}^{2}  \notag \\
&\leq &1-\frac{1}{4}(\mathrm{Tr}\sqrt{\rho _{AB}})^{2},  \label{diyibu}
\end{eqnarray}%
where we have used the Cauchy-Schwarz inequality $\sum_{k=1}^{4}\left\vert
U_{A}\otimes U_{B}\sqrt{\rho _{AB}}U_{A}^{\dagger }\otimes U_{B}^{\dagger
}\right\vert _{kk}^{2}\geq \frac{1}{4}(\mathrm{Tr}\sqrt{\rho _{AB}})^{2}$.
This inequality is saturated iff $U_{A}\otimes U_{B}\sqrt{\rho _{AB}}%
U_{A}^{\dagger }\otimes U_{B}^{\dagger }$ has identical diagonal entries.
This can be easily found based on the lemma presented in the Appendix; one
can first find a proper $U_{A}$ such that $R_{11}=R_{33}$ \ and $%
R_{22}=R_{44}$ with $R=U_{A}\otimes \mathbb{I}\sqrt{\rho _{AB}}%
U_{A}^{\dagger }\otimes \mathbb{I}$, and then find another matrix $U_{B}$
such that $\left[ \mathbb{I}\otimes U_{B}R\mathbb{I}\otimes U_{B}^{\dagger }%
\right] _{11}=\left[ \mathbb{I}\otimes U_{B}R\mathbb{I}\otimes
U_{B}^{\dagger }\right] _{22}$ and $\left[ \mathbb{I}\otimes U_{B}R\mathbb{I}%
\otimes U_{B}^{\dagger }\right] _{33}=\left[ \mathbb{I}\otimes U_{B}R\mathbb{%
I}\otimes U_{B}^{\dagger }\right] _{44}$.

If $\rho _{A}$ is degenerate but $\rho _{B}$ is not degenerate with spectral
decomposition $\rho _{B}=\sum_{k_{B}=0}^{1}\lambda _{k_{B}}\left\vert
k_{B}\right\rangle \left\langle k_{B}\right\vert $, then the projective
measurement that does not disturb $\rho _{A}$ and $\rho _{B}$ is $K=\left\{
K_{k_{A}k_{B}}=\left\vert k_{A}\right\rangle \left\langle k_{A}\right\vert
\otimes \left\vert k_{B}\right\rangle \left\langle k_{B}\right\vert \right\}
$; thus, only the basis vectors of subsystem $A$ need to be optimized.
Therefore, we can write $N(\rho _{AB})$ as
\begin{eqnarray}
N(\rho _{AB}) &=&\max_{\left\{ \left\vert k_{A}\right\rangle \right\}
,\left\{ \left\vert k_{B}\right\rangle \right\}
}\sum_{k_{A},k_{B}=0}^{1}I(\rho _{AB},K_{k_{A}k_{B}})  \notag \\
&=&1-\min_{\left\{ \left\vert k_{A}\right\rangle \right\}
}\sum_{k_{A},k_{B}=0}^{1}\left\vert \left\langle k_{A}\right\vert
\left\langle k_{B}\right\vert \sqrt{\rho _{AB}}\left\vert k_{A}\right\rangle
\left\vert k_{B}\right\rangle \right\vert ^{2}  \notag \\
&=&1-\min_{\left\{ \left\vert k_{A}\right\rangle \right\}
}\sum_{k_{A},k_{B}=0}^{1}\left\vert U_{A}\left\langle k_{B}\right\vert \sqrt{%
\rho _{AB}}\left\vert k_{B}\right\rangle U_{A}^{\dagger }\right\vert
_{k_{A}k_{A}}^{2}  \notag \\
&\leq &1-\frac{1}{2}\sum_{k_{B}=0}^{1}(\mathrm{Tr}\left( \left\langle
k\right\vert _{B}\sqrt{\rho _{AB}}\left\vert k\right\rangle _{B}\right)
)^{2},  \label{kexi}
\end{eqnarray}%
where we use the Cauchy-Schwarz inequality for each $\left\vert
k\right\rangle _{B}$. Again, the inequality can be saturated as seen from
the lemma given in the Appendix. A similar proof can also be achieved if $%
\rho _{A}$ is not degenerate but $\rho _{B}$ is degenerate. Thus, the proof
is complete.\hfill $\blacksquare $

\textbf{Theorem 2.-} For a bipartite "X"-type density matrix
\begin{equation}
\rho _{AB}=\left(
\begin{array}{cccc}
\rho _{11} & 0 & 0 & \rho _{14} \\
0 & \rho _{22} & \rho _{23} & 0 \\
0 & \rho _{23}^{\ast } & \rho _{33} & 0 \\
\rho _{14}^{\ast } & 0 & 0 & \rho _{44}%
\end{array}%
\right) ,  \label{RHO}
\end{equation}%
the SQD of $\rho _{AB}$ is given by%
\begin{eqnarray}
D(\rho _{AB}) &=&1-\frac{1}{4}[(Tr\sqrt{\left\vert \rho _{AB}\right\vert }%
)^{2}  \notag \\
&&+\max \{a_{0z}^{2}+a_{z0}^{2}+a_{zz}^{2},a_{xx}^{2},a_{yy}^{2}\}],
\label{dQIUJIE}
\end{eqnarray}%
where$\ $%
\begin{equation}
a_{ij}=Tr\left\{ \left( \sigma _{i}\otimes \sigma _{j}\right) \sqrt{%
\left\vert \rho _{AB}\right\vert }\right\} ,i,j=0,x,y,z  \label{abcdef}
\end{equation}%
with $\sigma _{0}$ and $\sigma _{x\backslash y\backslash z}$, denoting the
identity matrix and the corresponding Pauli matrices, respectively, and $%
\left\vert \cdot \right\vert $ representing the absolute values of the
matrix entries.

\textbf{Proof. }First, $\rho _{AB}$ can be rewritten as%
\begin{equation}
\rho _{AB}=\left(
\begin{array}{cccc}
\rho _{11} & 0 & 0 & \left\vert \rho _{14}\right\vert e^{i\theta } \\
0 & \rho _{22} & \left\vert \rho _{23}\right\vert e^{i\varphi } & 0 \\
0 & \left\vert \rho _{23}\right\vert e^{-i\varphi } & \rho _{33} & 0 \\
\left\vert \rho _{14}\right\vert e^{-i\theta } & 0 & 0 & \rho _{44}%
\end{array}%
\right) ;  \label{bianhuan}
\end{equation}%
\ hence, one can easily find local unitary transformations $U_{A}=diag(e^{-%
\frac{i\theta }{2}};e^{\frac{i\varphi }{2}})$ and $U_{B}=diag(e^{-\frac{%
i\theta }{2}};e^{-\frac{i\varphi }{2}})$ such that
\begin{equation}
\left\vert \rho _{AB}\right\vert =\left( U_{A}\otimes U_{B}\right) \rho
_{AB}\left( U_{A}\otimes U_{B}\right) ^{\dag }.
\end{equation}%
Thus, an "X"-type state $\rho _{AB}$ with 7 free real parameters has been
converted into an "X"-type state $\left\vert \rho _{AB}\right\vert $ in real
space with only 5 free parameters, but the SQDs of both are the same due to
the invariance under local unitary transformations. In this sense,
calculating the SQD of $\rho _{AB}$ is equivalent to evaluating the SQD of $%
\left\vert \rho _{AB}\right\vert $.

By substituting $\left\vert \rho _{AB}\right\vert $ into Definition 1, one
obtains%
\begin{eqnarray}
&&D(\left\vert \rho _{AB}\right\vert )  \notag \\
&=&1-\max_{\left\{ \left\vert k_{A}\right\rangle \right\} ,\left\{
\left\vert k_{B}\right\rangle \right\} }\sum_{k_{A},k_{B}=0}^{1}\left\vert
\left\langle k_{A}\right\vert \left\langle k_{B}\right\vert \sqrt{\left\vert
\rho _{AB}\right\vert }\left\vert k_{A}\right\rangle \left\vert
k_{B}\right\rangle \right\vert ^{2}  \notag \\
&=&1-\max_{U_{A},U_{B}}\sum_{k=1}^{4}\left\vert U_{A}\otimes U_{B}\sqrt{%
\left\vert \rho _{AB}\right\vert }U_{A}^{\dag }\otimes U_{B}^{\dagger
}\right\vert _{kk}^{2}.  \label{DD1}
\end{eqnarray}%
Obviously, the remaining task is to find an optimal $U=U_{A}(\theta
_{1},\varphi _{1})\otimes U_{B}(\theta _{2},\varphi _{2})$ with $%
U_{A}(\theta _{1},\varphi _{1})=\left(
\begin{array}{cc}
\cos \theta _{1} & e^{i\varphi _{1}}\sin \theta _{1} \\
-e^{-i\varphi _{1}}\sin \theta _{1} & \cos \theta _{1}%
\end{array}%
\right) $ and $U_{B}(\theta _{2},\varphi _{2})=\left(
\begin{array}{cc}
\cos \theta _{2} & e^{i\varphi _{2}}\sin \theta _{2} \\
-e^{-i\varphi _{2}}\sin \theta _{2} & \cos \theta _{2}%
\end{array}%
\right) $ such that the objective function
\begin{equation}
\tilde{J}(U)=\sum_{k=1}^{4}\left\vert U\sqrt{\left\vert \rho
_{AB}\right\vert }U^{\dagger }\right\vert _{kk}^{2}  \label{QU}
\end{equation}%
is maximized. For convenience, let $G=\sqrt{\left\vert \rho _{AB}\right\vert
}$. Since $\left\vert \rho _{AB}\right\vert $ is an "X"-type state in real
space, $\sqrt{\left\vert \rho _{AB}\right\vert }$ is also an "X"-type matrix
in real space. Then, let us suppose that the unitary transformation $U$ for
updating the matrix $\sqrt{\left\vert \rho _{AB}\right\vert }$ satisfies $%
G^{\prime }=U\sqrt{\left\vert \rho _{AB}\right\vert }U^{\dagger }$, where
the entries of $G^{^{\prime }}$ are denoted by $G_{ij}^{\prime }$, $%
i,j=1,2,3,4$; then, the optimization problem on $\tilde{J}(U)$ in Equation (%
\ref{QU}) is equivalent to the problem%
\begin{eqnarray}
&&\max_{\{\theta _{1},\varphi _{1},\theta _{2},\varphi _{2}\}}\tilde{J}%
(\theta _{1},\varphi _{1},\theta _{2},\varphi _{2})  \notag \\
&=&\max_{\{\theta _{1},\varphi _{1},\theta _{2},\varphi _{2}\}}\{\left\vert
G_{11}^{\prime }\right\vert ^{2}+\left\vert G_{22}^{\prime }\right\vert
^{2}+\left\vert G_{33}^{\prime }\right\vert ^{2}+\left\vert G_{44}^{\prime
}\right\vert ^{2}\}  \notag \\
&=&\max_{\{\theta _{1},\varphi _{1},\theta _{2},\varphi _{2}\}}\frac{1}{4}%
[\left\vert G_{11}^{\prime }+G_{22}^{\prime }-G_{33}^{\prime
}-G_{44}^{\prime }\right\vert ^{2}  \notag \\
&&+\left\vert G_{11}^{\prime }-G_{22}^{\prime }+G_{33}^{\prime
}-G_{44}^{\prime }\right\vert ^{2}  \notag \\
&&+\left\vert G_{11}^{\prime }-G_{22}^{\prime }-G_{33}^{\prime
}+G_{44}^{\prime }\right\vert ^{2}  \notag \\
&&+\left\vert G_{11}^{\prime }+G_{22}^{\prime }+G_{33}^{\prime
}+G_{44}^{\prime }\right\vert ^{2}].  \label{sixiang}
\end{eqnarray}%
Since the trace of a matrix is preserved under a unitary transformation, we
have $\sum_{i}G_{ii}^{\prime }=\sum_{i}G_{ii}=Tr\sqrt{\left\vert \rho
_{AB}\right\vert }$. Thus,
\begin{eqnarray}
D(\rho _{AB}) &=&1-\frac{1}{4}[(Tr\sqrt{\left\vert \rho _{AB}\right\vert }%
)^{2}  \notag \\
&&+\max_{\{\theta _{1},\varphi _{1},\theta _{2},\varphi _{2}\}}J(\theta
_{1},\varphi _{1},\theta _{2},\varphi _{2})],  \label{rewrite}
\end{eqnarray}%
where%
\begin{eqnarray}
&&J(\theta _{1},\varphi _{1},\theta _{2},\varphi _{2})  \notag \\
&=&\left\{ \left\vert G_{11}^{\prime }+G_{22}^{\prime }-G_{33}^{\prime
}-G_{44}^{\prime }\right\vert ^{2}\right.   \notag \\
&&+\left\vert G_{11}^{\prime }-G_{22}^{\prime }+G_{33}^{\prime
}-G_{44}^{\prime }\right\vert ^{2}  \notag \\
&&+\left. \left\vert G_{11}^{\prime }-G_{22}^{\prime }-G_{33}^{\prime
}+G_{44}^{\prime }\right\vert ^{2}\right\} .  \label{sanxiang}
\end{eqnarray}%
In addition, the optimization in Equation (\ref{rewrite}) can be rewritten as%
\begin{gather}
\max_{\{\theta _{1},\varphi _{1},\theta _{2},\varphi _{2}\}}J(\theta
_{1},\varphi _{1},\theta _{2},\varphi _{2})  \notag \\
=\max_{\mathbf{x,y}}\left\{ \mathbf{x}^{\intercal }\left( \mathbf{AA}%
^{\dagger }+C\mathbf{yy}^{\intercal }C^{\dagger }\right) \mathbf{x}+\mathbf{y%
}^{\intercal }\mathbf{BB}^{\dagger }\mathbf{y}\right\} ,  \label{liangxiang}
\end{gather}%
where $\mathbf{A}=(%
\begin{array}{ccc}
a & 0 & 0%
\end{array}%
)^{\intercal }$, $\mathbf{B}=(%
\begin{array}{ccc}
b & 0 & 0%
\end{array}%
)^{\intercal }$, $C=\left(
\begin{array}{ccc}
c & 0 & 0 \\
0 & d & 0 \\
0 & 0 & e%
\end{array}%
\right) $, $\mathbf{x}=(%
\begin{array}{ccc}
\cos 2\theta _{1} & -\sin 2\theta _{1}\cos \varphi _{1} & -\sin 2\theta
_{1}\sin \varphi _{1}%
\end{array}%
)^{\intercal }$, and\textbf{\ }$\mathbf{y}=(%
\begin{array}{ccc}
\cos 2\theta _{2} & -\sin 2\theta _{2}\cos \varphi _{2} & -\sin 2\theta
_{2}\sin \varphi _{2}%
\end{array}%
)^{\intercal }$. Notably, the real parameters $a=a_{z0}$, $b=a_{0z}$, $%
c=a_{zz}$, $d=a_{xx}$ and $e=a_{yy}$ are defined by Equation (\ref{abcdef}).
The one-to-one correspondence between Equation (\ref{sanxiang}) and Equation
(\ref{liangxiang}) is as follows:%
\begin{eqnarray}
\left\vert G_{11}^{\prime }+G_{22}^{\prime }-G_{33}^{\prime }-G_{44}^{\prime
}\right\vert ^{2} &=&\mathbf{x}^{\intercal }\mathbf{AA}^{\dagger }\mathbf{x,}
\notag \\
\left\vert G_{11}^{\prime }-G_{22}^{\prime }+G_{33}^{\prime }-G_{44}^{\prime
}\right\vert ^{2} &=&\mathbf{y}^{\intercal }\mathbf{BB}^{\dagger }\mathbf{y,}
\notag \\
\left\vert G_{11}^{\prime }-G_{22}^{\prime }-G_{33}^{\prime }+G_{44}^{\prime
}\right\vert ^{2} &=&\mathbf{x}^{\intercal }C\mathbf{yy}^{\intercal
}C^{\dagger }\mathbf{x.}  \label{jieshi}
\end{eqnarray}

The first term in Equation (\ref{liangxiang}) is the well-known Rayleigh
quotient; \cite{ruili} thus, it can be maximized by the maximal eigenvalue $%
\lambda _{\max }$ of the rank-2 matrix $\mathbf{AA}^{\dagger }+C\mathbf{yy}%
^{\intercal }C^{\dagger }$, which can be found through a simple algebraic
derivation as follows:%
\begin{eqnarray}
\lambda _{\max }(\theta _{2},\varphi _{2}) &=&\frac{1}{2}\{a^{2}+c^{2}\cos
^{2}2\theta _{2}+d^{2}\sin ^{2}2\theta _{2}\cos ^{2}\varphi _{2}  \notag \\
&&+[(a^{2}+d^{2}\sin ^{2}2\theta _{2}\cos ^{2}\varphi _{2}  \notag \\
&&+c^{2}\cos ^{2}2\theta _{2}+e^{2}\sin ^{2}2\theta _{2}\sin ^{2}\varphi
_{2})^{2}  \notag \\
&&-4a^{2}\sin ^{2}2\theta _{2}(d^{2}\cos ^{2}\varphi _{2}+e^{2}\sin
^{2}\varphi _{2})]^{\frac{1}{2}}  \notag \\
&&+e^{2}\sin ^{2}2\theta _{2}\sin ^{2}\varphi _{2}\}.  \label{yixiang}
\end{eqnarray}%
Correspondingly, the optimization defined in Equation (\ref{liangxiang}) can
be converted into
\begin{equation}
\max_{\{\theta _{2},\varphi _{2}\}}J(\theta _{2},\varphi _{2})=\lambda
_{\max }(\theta _{2},\varphi _{2})+2b^{2}\cos ^{2}2\theta _{2}.  \label{yfi}
\end{equation}%
Let $r=\cos ^{2}2\theta _{2}$, $s=\sin ^{2}2\theta _{2}\cos ^{2}\varphi _{2}$%
, $t=\sin ^{2}2\theta _{2}\sin ^{2}\varphi _{2}$, $%
M=a^{2}+c^{2}r+d^{2}s+e^{2}t$, and $N=4a^{2}(d^{2}s+e^{2}t)$; then, Equation
(\ref{yfi}) can be simplified to

\begin{eqnarray}
\max_{\{r,s,t\}}J(r,s,t) &=&\frac{1}{2}[a^{2}+(2b^{2}+c^{2})r  \notag \\
&&+d^{2}s+e^{2}t+\sqrt{M^{2}-N}]  \notag \\
\text{s.t. }r,s,t &\geq &0,r+s+t=1.  \label{fop}
\end{eqnarray}

\textit{Case 1:} $a=0$.

In this case, the optimized function $J(r,s,t)$ in Equation (\ref{fop})
becomes%
\begin{equation}
J(r,s,t)=(b^{2}+c^{2})r+d^{2}s+e^{2}t.  \label{a=0}
\end{equation}%
It is obvious that $\max J=\max \{b^{2}+c^{2},d^{2},e^{2}\}$;\ which can be
found from Equation (\ref{dQIUJIE}).

\textit{Case 2:} $c=0$.

In this case, $J(r,s,t)$ in Equation (\ref{fop}) becomes a piecewise
function as follwos:%
\begin{equation}
J(r,s,t)=\left\{
\begin{array}{cc}
b^{2}r+a^{2} & d^{2}s+e^{2}t\leq a^{2} \\
b^{2}r+d^{2}s+e^{2}t & d^{2}s+e^{2}t>a^{2}%
\end{array}%
\right. .  \label{c=0}
\end{equation}%
If $d^{2}s+e^{2}t\leq a^{2}$, then the maximum value of $b^{2}r+a^{2}$ is $%
a^{2}+b^{2}$ . If $d^{2}s+e^{2}t>a^{2}$, then the maximum value of $%
b^{2}r+d^{2}s+e^{2}t$ is $\max \{b^{2},d^{2},e^{2}\}$. Combining these two
results, one can immediately find that the maximum value of $J(r,s,t)$
throughout the whole range is given by $\max J=\max
\{a^{2}+b^{2},d^{2},e^{2}\}$, which again corresponds to Equation (\ref%
{dQIUJIE}).

\textit{Case 3: }$a\neq 0,c\neq 0,d^{2}=e^{2}$.

In this case, we can rewrite $J(r,s,t)$ as
\begin{eqnarray}
J(r) &=&\frac{1}{2}\{a^{2}+d^{2}+(2b^{2}+c^{2}-d^{2})r  \notag \\
&&+[(a^{2}+d^{2}+c^{2}r-d^{2}r)^{2}  \notag \\
&&-4a^{2}d^{2}(1-r)]^{\frac{1}{2}}\}.  \label{ddengyue}
\end{eqnarray}%
Only one parameter, $r$, remains. Let us solve for it in the following two
cases.

\textit{Case 3.1:} $d^{2}\leq b^{2}+c^{2}$. We can easily find that%
\begin{eqnarray}
J &\leq &\max \frac{1}{2}[a^{2}+d^{2}+(2b^{2}+c^{2}-d^{2})r  \notag \\
&&+\sqrt{(a^{2}+d^{2}+c^{2}r-d^{2}r)^{2}}]  \notag \\
&=&\max \{a^{2}+d^{2}+(b^{2}+c^{2}-d^{2})r\}  \notag \\
&=&a^{2}+b^{2}+c^{2}.  \label{Jmax}
\end{eqnarray}%
Equation (\ref{Jmax}) will be saturated if $r=1$; this case is included in
Equation (\ref{dQIUJIE}).

\textit{Case 3.2:} $d^{2}>b^{2}+c^{2}$. In this case, we would like to
rewrite $J(r)$ in Equation (\ref{ddengyue}) as
\begin{equation}
J(X)=\frac{1}{2}\{K+QX+\sqrt{X^{2}-D}\},  \label{djianhua}
\end{equation}%
where $K=a^{2}+d^{2}-Q(a^{2}+d^{2}+\frac{2a^{2}d^{2}}{c^{2}-d^{2}})$, $Q=%
\frac{2b^{2}+c^{2}-d^{2}}{c^{2}-d^{2}}$, $X=a^{2}+d^{2}+\frac{2a^{2}d^{2}}{%
c^{2}-d^{2}}+(c^{2}-d^{2})r$ and $D=\frac{4a^{2}d^{2}}{c^{2}-d^{2}}(\frac{%
a^{2}d^{2}}{c^{2}-d^{2}}+a^{2}+c^{2})$.

First, we would like to consider the equation $X^{2}-D=0$, which implies the
sufficient and necessary conditions $r=0$ and $a^{2}=d^{2}$. One can find
that $J(r=0)=\max \{a^{2},d^{2}\}\leq \max \{J(r=1),J(s=1),J(t=1)\}$, which
is given by Equation (\ref{dQIUJIE}).

Now, let us consider the case of $r\neq 0$, which implies that $X^{2}-D\neq
0 $. Based on the Lagrangian multiplier method, the vanishing derivative of $%
J(X)$ with respect to the variable $X$ reads%
\begin{equation}
\frac{\partial J}{\partial X}=\frac{1}{2}\{Q+\frac{X}{\sqrt{X^{2}-D}}\}=0,
\label{yijeidaoshu}
\end{equation}%
or
\begin{equation}
(Q^{2}-1)X^{2}=Q^{2}D.  \label{haode}
\end{equation}

\textit{Case 3.2.1:} $b=0$. We have\textit{\ }$Q=1$. In this case, Equation (%
\ref{haode}) is valid iff $D=0$. The equation $D=0$ implies that%
\begin{equation}
\frac{a^{2}d^{2}}{c^{2}-d^{2}}+a^{2}+c^{2}=0,  \label{DD}
\end{equation}%
which further leads to $d^{2}=a^{2}+c^{2}$. That is, if $d^{2}\neq
a^{2}+c^{2}$, then the derivative of $J(X)$ in Equation (\ref{yijeidaoshu})
can not be zero. This means that the function $J(X)$ has no extreme point,
and consequently, its maximum value lies at the boundary; this situation,
corresponds to\ Equation (\ref{dQIUJIE}). If $d^{2}=a^{2}+c^{2}$, then the
function $J(r)$ in Equation (\ref{ddengyue}) can be rewritten as $%
J(r)=d^{2}=a^{2}+b^{2}+c^{2}$, which obviously includes the endpoints of $X$
(i.e., $r$).

\textit{Case 3.2.2:} $b\neq 0$. We have\textit{\ }$Q^{2}\neq 1$. Thus, from
Equation (\ref{haode}), we arrive at
\begin{equation}
X^{2}=\frac{Q^{2}D}{(Q^{2}-1)}.  \label{yijie}
\end{equation}%
Taking the second derivative of $J(X)$ with respect to $X$ leads to%
\begin{equation}
\frac{\partial ^{2}J}{\partial X^{2}}=\frac{-D}{\sqrt{(X^{2}-D)^{3}}}.
\label{2jiedoa}
\end{equation}%
In order to find a local maximum value of \ $J(X)$ as defined in Equation (%
\ref{djianhua}), we would expect the existence of a point at which $\frac{%
\partial ^{2}J}{\partial X^{2}}<0$, i.e., $D>0$. Considering the validity of
Equation (\ref{yijie}) for $D>0$, one can obtain $Q^{2}>1$, which implies
that $d^{2}<b^{2}+c^{2}$. This contradicts $d^{2}>b^{2}+c^{2}$, as claimed
in \textit{Case 3.2}. Thus, $J(X)$ has no local maximum value subject to
Equation (\ref{yijie}). In other words, the maximum value must lie at the
endpoints of $X$. This shows that $\max J(X)=\max \{a^{2}+b^{2}+c^{2},d^{2}\}
$, which is included in Equation (\ref{dQIUJIE}).

\textit{Case 4:} $a\neq 0,c\neq 0,d^{2}\neq e^{2}$.

Consider the Lagrangian function for $J(r,s,t)$ in Equation (\ref{fop}),
i.e.,%
\begin{eqnarray}
L(r,s,t,\lambda ) &=&\frac{1}{2}[a^{2}+c^{2}r+d^{2}s+e^{2}t+\sqrt{M^{2}-N}]
\notag \\
&&+b^{2}r+\lambda (r+s+t-1)  \label{lagelangri}
\end{eqnarray}%
where $\lambda $ is the Lagrangian multiplier. First, one can easily find
that the derivative of $L(r,s,t,\lambda )$ tends toward infinity at the
point $T(r=0,s=\frac{a^{2}-e^{2}}{d^{2}-e^{2}},t=\frac{d^{2}-a^{2}}{%
d^{2}-e^{2}})$, where $M^{2}=N$. Since $r$, $s$, and $t$ are not negative,
one can find that the point $T$ makes sense only for $\min
\{d^{2},e^{2}\}<a^{2}<\max \{d^{2},e^{2}\}$ and $M=2a^{2}$, from which we
can find that $J(T)=a^{2}$. It is obvious that $J(T)<J(1,0,0)$. Therefore, $%
J(T)$ is not the maximum value of $J(r,s,t)$, and the point $T$ can be
safely neglected.

Now, let us consider the function $L(r,s,t,\lambda )$ excluding the point $T$
(that is, $M^{2}\neq N$ or $M\neq 2a^{2}$). Based on the Lagrangian
multiplier method, the derivatives with respect to the parameters of $%
L(r,s,t,\lambda )$ are given by%
\begin{equation}
\left\{
\begin{array}{c}
\frac{\partial L}{\partial r}=\frac{1}{2}c^{2}+b^{2}+\lambda +\frac{Mc^{2}}{2%
\sqrt{M^{2}-N}}=0 \\
\frac{\partial L}{\partial s}=\frac{1}{2}d^{2}+\lambda +\frac{M-2a^{2}}{2%
\sqrt{M^{2}-N}}d^{2}=0 \\
\frac{\partial L}{\partial t}=\frac{1}{2}e^{2}+\lambda +\frac{M-2a^{2}}{2%
\sqrt{M^{2}-N}}e^{2}=0%
\end{array}%
\right. .  \label{qiudao1}
\end{equation}%
From this equation array, one can directly arrive at%
\begin{equation}
\left\{
\begin{array}{c}
c^{2}+2b^{2}-d^{2}=\frac{Md^{2}-Mc^{2}-2a^{2}d^{2}}{\sqrt{M^{2}-N}} \\
c^{2}+2b^{2}-e^{2}=\frac{Me^{2}-Mc^{2}-2a^{2}e^{2}}{\sqrt{M^{2}-N}}%
\end{array}%
\right.  \label{MNshi1}
\end{equation}%
which further leads to%
\begin{equation}
d^{2}-e^{2}=\frac{(d^{2}-e^{2})(2a^{2}-M)}{\sqrt{M^{2}-N}}.  \label{a1}
\end{equation}%
Since $d^{2}\neq e^{2}$, Equation (\ref{a1}) implies that
\begin{eqnarray}
N &=&4a^{2}(M-a^{2}),  \label{nn1} \\
M &<&2a^{2}.  \label{nn2}
\end{eqnarray}%
By substituting Equation (\ref{nn1}) into Equation (\ref{MNshi1}), one can
obtain%
\begin{equation}
b^{2}M^{2}-a^{2}(c^{2}+4b^{2})M+2a^{4}(c^{2}+2b^{2})=0.  \label{MNshi2}
\end{equation}

\textit{Case 4.1:} $b\neq 0$. One solution to Equation (\ref{MNshi2}) is $%
M=2a^{2} $, which lies just at the excluded point $T$, and the other
solution is $M=2a^{2}+\frac{a^{2}c^{2}}{b^{2}}$, which obviously contradicts
Equation (\ref{nn2}). Consequently, no valid solution is obtained, namely,
no extreme value exists in this case.

\textit{Case 4.2:} $b=0$. The unique solution to Equation (\ref{MNshi2}) is $%
M=2a^{2}$, which is again invalid. Therefore, in this case, the maximum
value of $J(r,s,t)$ can only be attained at the boundary.

Now, we will check the maximum value at the boundary. For $r=0$, one can
easily find that $\max J(r,s,t)=\max \{a^{2},d^{2},e^{2}\}$ $\leq \max
\{a^{2}+b^{2}+c^{2},d^{2},e^{2}\}$, which is given by Equation (\ref{dQIUJIE}%
). For $s=0$ or $t=0$, the objective function in the optimization question
becomes the same as that in Equation (\ref{ddengyue}), up to the possible
replacement of $d^{2}$ with $e^{2}$. Following the same procedures in
\textit{Case 3.1} and \textit{Case 3.2}, one can again find that the maximum
values are always obtained by Equation (\ref{dQIUJIE}). Finally, one can
note that Equation (\ref{dQIUJIE}) simply gives the values of the boundary
points with $r=1$, or $s=1$ or $t=1$.

In summary, we have shown that in any arbitrary case, the maximum value is
given by Equation (\ref{dQIUJIE}), which completes the proof. \hfill $%
\blacksquare $

\textbf{Theorem 3.-} If a bipartite quantum state $\rho _{AB}$ is block
diagonal, i.e., $\rho _{AB}=\left(
\begin{array}{cc}
\rho _{11} & \rho _{12} \\
\rho _{12}^{\ast } & \rho _{22}%
\end{array}%
\right) \oplus \left(
\begin{array}{cc}
\rho _{33} & \rho _{34} \\
\rho _{34}^{\ast } & \rho _{44}%
\end{array}%
\right) $, or equivalently, $\rho _{AB}^{\prime }=U_{swap}\rho
_{AB}U_{swap}^{\dag }$ where $U_{swap}$ denotes the bipartite swapping
operation, then the SQD of $\rho _{AB}$ ($\rho _{AB}^{\prime }$) is given by%
\begin{eqnarray}
D(\rho _{AB}) &=&1-\frac{1}{4}[(\text{Tr}\sqrt{\rho _{AB}})^{2}  \notag \\
&&+a_{z0}^{2}+\lambda _{\max }(\mathbf{BB}^{\dagger }+\mathbf{CC}^{\dag })],
\label{discord3}
\end{eqnarray}%
where $\mathbf{B=(}%
\begin{array}{ccc}
a_{0x} & a_{0y} & a_{0z}%
\end{array}%
\mathbf{)^{\intercal }}$ and $\mathbf{C=}(%
\begin{array}{ccc}
a_{zx} & a_{zy} & a_{zz}%
\end{array}%
)^{\intercal }$, with $\left\vert \cdot \right\vert $ and $a_{ij}$ defined
in the same way as in Theorem 2.

\textbf{Proof. }Based on the definition of the SQD given in Equation (\ref%
{def1}), we have
\begin{equation}
D(\rho _{AB})=1-\max_{U_{A},U_{B}}\sum_{k=1}^{4}\left\vert U_{AB}\sqrt{\rho
_{AB}}U_{AB}^{\dag }\right\vert _{kk}^{2}  \label{bdd}
\end{equation}%
where $U_{AB}=U_{A}(\theta _{1},\varphi _{1})\otimes U_{B}(\theta
_{2},\varphi _{2})$. Following the same procedure applied to proceed from
Equation (\ref{DD1}) to Equation (\ref{sixiang}), one can rewrite Equation (%
\ref{bdd}) as
\begin{gather}
D(\rho _{AB})=1-  \notag \\
\frac{1}{4}\left[ (\mathrm{Tr}\sqrt{\rho _{AB}})^{2}+\max_{\theta
_{1},\varphi _{1},\theta _{2},\varphi _{2}}J(\theta _{1},\varphi _{1},\theta
_{2},\varphi _{2})\right] .
\end{gather}%
Here,%
\begin{eqnarray}
&&\max_{\theta _{1},\varphi _{1},\theta _{2},\varphi _{2}}J(\theta
_{1},\varphi _{1},\theta _{2},\varphi _{2})  \notag \\
&=&\max_{\mathbf{x},\mathbf{y}}\left\{ \mathbf{x}^{\intercal }\left( \mathbf{%
AA}^{\dagger }+C\mathbf{yy}^{\intercal }C^{\dagger }\right) \mathbf{x}+%
\mathbf{y}^{\intercal }\mathbf{B}^{\prime }\mathbf{B}^{\prime \dagger }%
\mathbf{y}\right\} ,  \label{xinliangxiang}
\end{eqnarray}%
where $\mathbf{A}=(%
\begin{array}{ccc}
a_{z0} & 0 & 0%
\end{array}%
)^{\intercal }$, $\mathbf{B}^{\prime }=(%
\begin{array}{ccc}
a_{0z} & a_{0x} & -a_{0y}%
\end{array}%
)^{\intercal }$, $C=\left(
\begin{array}{ccc}
a_{zz} & a_{zx} & a_{zy} \\
0 & 0 & 0 \\
0 & 0 & 0%
\end{array}%
\right) $, $\mathbf{x}=(%
\begin{array}{ccc}
\cos 2\theta _{1} & -\sin 2\theta _{1}\cos \varphi _{1} & -\sin 2\theta
_{1}\sin \varphi _{1}%
\end{array}%
)^{\intercal }$, and $\mathbf{y}=(%
\begin{array}{ccc}
\cos 2\theta _{2} & -\sin 2\theta _{2}\cos \varphi _{2} & -\sin 2\theta
_{2}\sin \varphi _{2}%
\end{array}%
)^{\intercal }$ with $a_{ij}=Tr\left\{ \left( \sigma _{i}\otimes \sigma
_{j}\right) \sqrt{\left\vert \rho _{AB}\right\vert }\right\} $. It is
obvious that the rank of the matrix $\mathbf{AA}^{\dagger }+C\mathbf{yy}%
^{\intercal }C^{\dagger }$ is 1; therefore, the maximal eigenvalue of $%
\mathbf{AA}^{\dagger }+C\mathbf{yy}^{\intercal }C^{\dagger }$ is achieved
with the optimal $\mathbf{x}_{0}=(%
\begin{array}{ccc}
1 & 0 & 0%
\end{array}%
)^{\intercal }$. Thus, the optimization problem presented in Equation (\ref%
{xinliangxiang}) becomes%
\begin{equation}
\max_{\theta _{2},\varphi _{2}}J(\theta _{2},\varphi _{2})=a_{z0}^{2}+\max_{%
\mathbf{y}}\mathbf{y}^{\intercal }(\mathbf{B}^{\prime }\mathbf{B}^{\prime
\dagger }+C^{\intercal }\mathbf{x}_{0}\mathbf{x}_{0}^{\intercal }C)\mathbf{y}%
.  \label{J}
\end{equation}%
The maximum value can be obtained with the maximal eigenvalue of the matrix $%
\mathbf{B}^{\prime }\mathbf{B}^{\prime \dagger }+C^{\intercal }\mathbf{x}_{0}%
\mathbf{x}_{0}^{\intercal }C$. Thus, one can easily find that the maximum
value of $J$ is given by%
\begin{equation}
\max_{\theta _{1},\varphi _{1},\theta _{2},\varphi _{2}}J(\theta
_{1},\varphi _{1},\theta _{2},\varphi _{2})=a_{z0}^{2}+\lambda _{\max }(%
\mathbf{BB}^{\dagger }+\mathbf{CC}^{\dag })
\end{equation}%
where $\mathbf{B=(}%
\begin{array}{ccc}
a_{0x} & -a_{0y} & a_{0z}%
\end{array}%
\mathbf{)^{\intercal }}$ and $\mathbf{C=}(%
\begin{array}{ccc}
a_{zx} & a_{zy} & a_{zz}%
\end{array}%
)^{\intercal }$. Thus the proof is complete.\hfill $\blacksquare $

\section{Applications}

To further demonstrate the validity of our three theorems, we will compare
our analytical expressions for the quantum correlations with the results
obtained numerically.

\textit{Example 1.- The\textit{\ }SMIN of $(2\otimes 2)$-dimensional general
quantum states.}

To demonstrate the validity of our Theorem 1, we consider a state $\rho _{G}$
formed by mixing a maximally mixed state and a randomly generated $(2\otimes
2)$-dimensional mixed state, expressed as $G=G_{1}+iG_{2}$, where%
\begin{equation}
G_{1}=\left(
\begin{array}{cccc}
0.2409 & 0.1612 & -0.0787 & 0.1945 \\
0.1612 & 0.3006 & -0.1008 & 0.1707 \\
-0.0787 & -0.1008 & 0.1899 & -0.0732 \\
0.1945 & 0.1707 & -0.0732 & 0.2686%
\end{array}%
\right) ,
\end{equation}%
\begin{equation}
G_{2}=\left(
\begin{array}{cccc}
0 & -0.0551 & -0.0779 & 0.0362 \\
0.0551 & 0 & -0.1395 & 0.0742 \\
0.0779 & 0.1395 & 0 & 0.1295 \\
-0.0362 & -0.0742 & -0.1295 & 0%
\end{array}%
\right) ,
\end{equation}%
with%
\begin{equation}
\rho _{G}=\frac{1-x}{4}\mathbb{I}_{4}+xG,x\in \lbrack 0,1].  \label{G}
\end{equation}%
The SMIN results are plotted versus $x$ in Figure 1. The solid line
corresponds to the analytical results given by Theorem 1 and the points
marked with "+" symbols denote the numerical results. It is obvious that the
analytical and numerical results are consistent with each other.
\begin{figure}[tbp]
\centering
\includegraphics[width=1\columnwidth,height=1.8in]{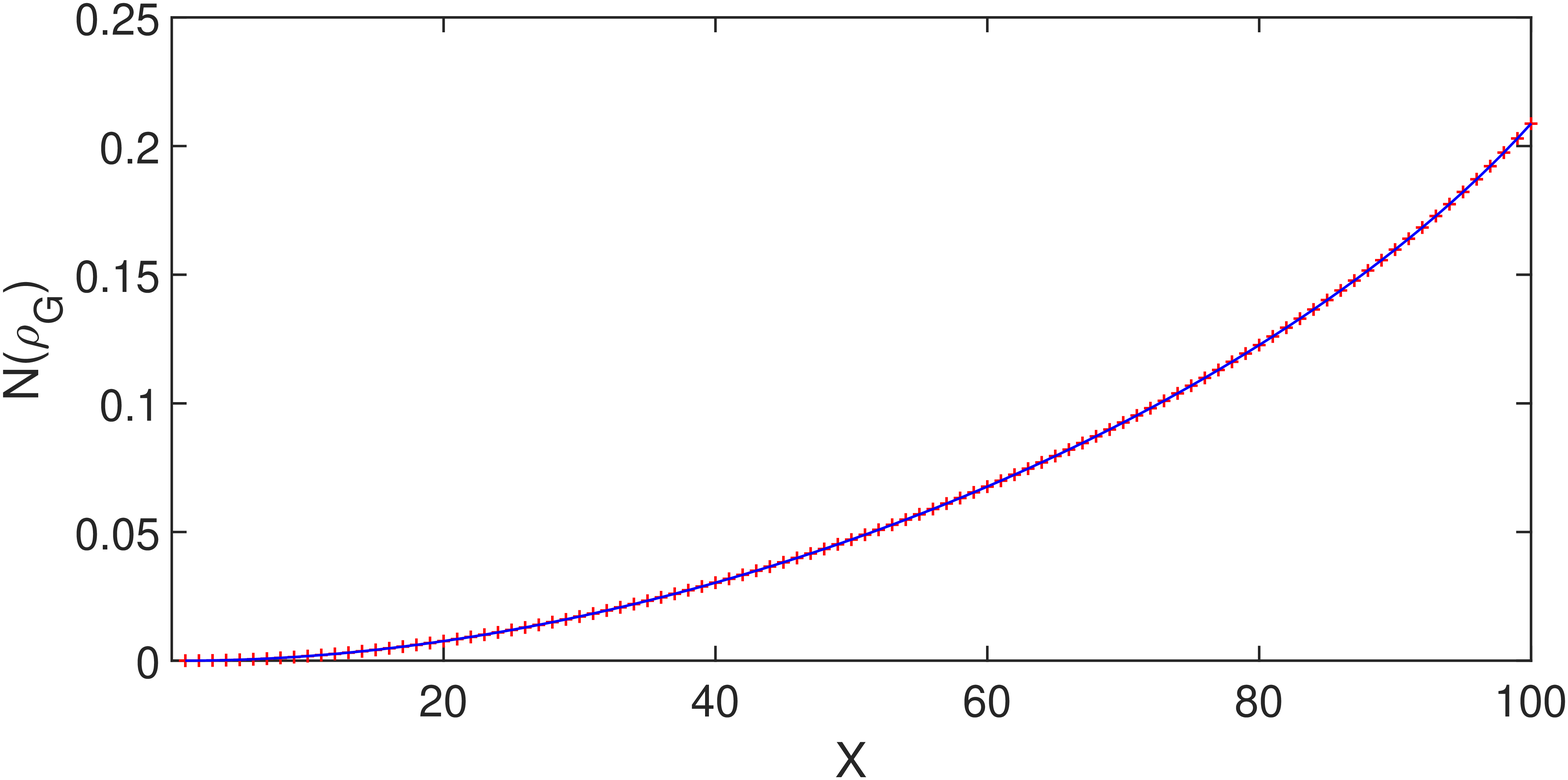}
\caption{(color online) SMIN results for randomly generated states $\protect%
\rho _{G}$ vs. $x$. The solid line corresponds to the strictly analytical
expressions of Equation (\protect\ref{nd}), (\protect\ref{MINjiexi}) and (%
\protect\ref{M1}), while the numerical solutions are marked with "+"
symbols.}
\end{figure}

\textit{Example 2.- The SQD of $(2\otimes 2)$-dimensional Werner states.}

To demonstrate the validity of Theorem 2, we consider the Werner states.
Werner states are the "X"-type states expressed as%
\begin{equation}
\rho _{W}=\frac{2-x}{6}\mathbb{I}_{4}+\frac{2x-1}{6}V,x\in \lbrack -1,1],
\label{W}
\end{equation}%
where $V=\sum_{kl}\left\vert kl\right\rangle \left\langle lk\right\vert $
denotes the swap operator. On the basis of our definition, the SQD $D(\rho
_{W})$ can be calculated as%
\begin{equation}
D(\rho _{AB})=\frac{1}{6}(2-x-\sqrt{3(1-x^{2})}).  \label{W1}
\end{equation}%
For comparison, we plot both the analytical expression given in Equation (%
\ref{W1}) and the numerical results in Figure 2. Again, the analytical and
numerical results show complete consistency.
\begin{figure}[tbp]
\centering
\includegraphics[width=1\columnwidth,height=1.8in]{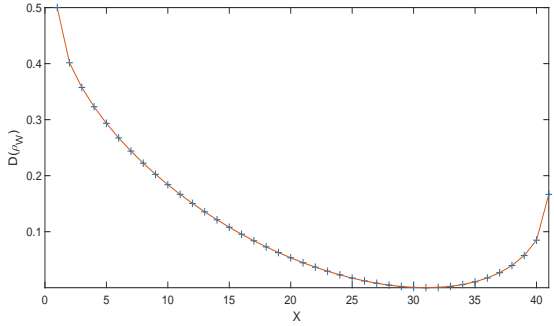}
\caption{(color online) SQD results for Werner states vs. $x$. The solid
line and the "+" symbols correspond to the strictly analytical expression of
Equation (\protect\ref{W1}) and the numerical results, respectively. }
\end{figure}

\textit{Example 3.- The SQD of $(2\otimes 2)$-dimensional general quantum
states}

To further demonstrate the validity of our Theorem 2, we consider a general
state, denoted by $\rho _{R}$, that has the same form as the state $\rho
_{G} $ in Equation (\ref{G}) but the matrix $G$ replaced by $R$, where the
matrix $R=R_{1}+iR_{2}$, with
\begin{equation}
R_{1}=\left(
\begin{array}{cccc}
0.2481 & 0 & 0 & 0.0103 \\
0 & 0.2083 & 0.0285 & 0 \\
0 & 0.0285 & 0.4657 & 0 \\
0.0103 & 0 & 0 & 0.0779%
\end{array}%
\right) ,
\end{equation}%
\begin{equation}
R_{2}=\left(
\begin{array}{cccc}
0 & 0 & 0 & -0.0141 \\
0 & 0 & 0.0877 & 0 \\
0 & -0.0877 & 0 & 0 \\
0.0141 & 0 & 0 & 0%
\end{array}%
\right) ,
\end{equation}%
is a randomly generated "X"-type state. We compare our analytical results
for the SQD of such states with the numerical results in Figure 3 which
again shows perfect consistency.
\begin{figure}[tbp]
\centering
\includegraphics[width=1\columnwidth,height=1.8in]{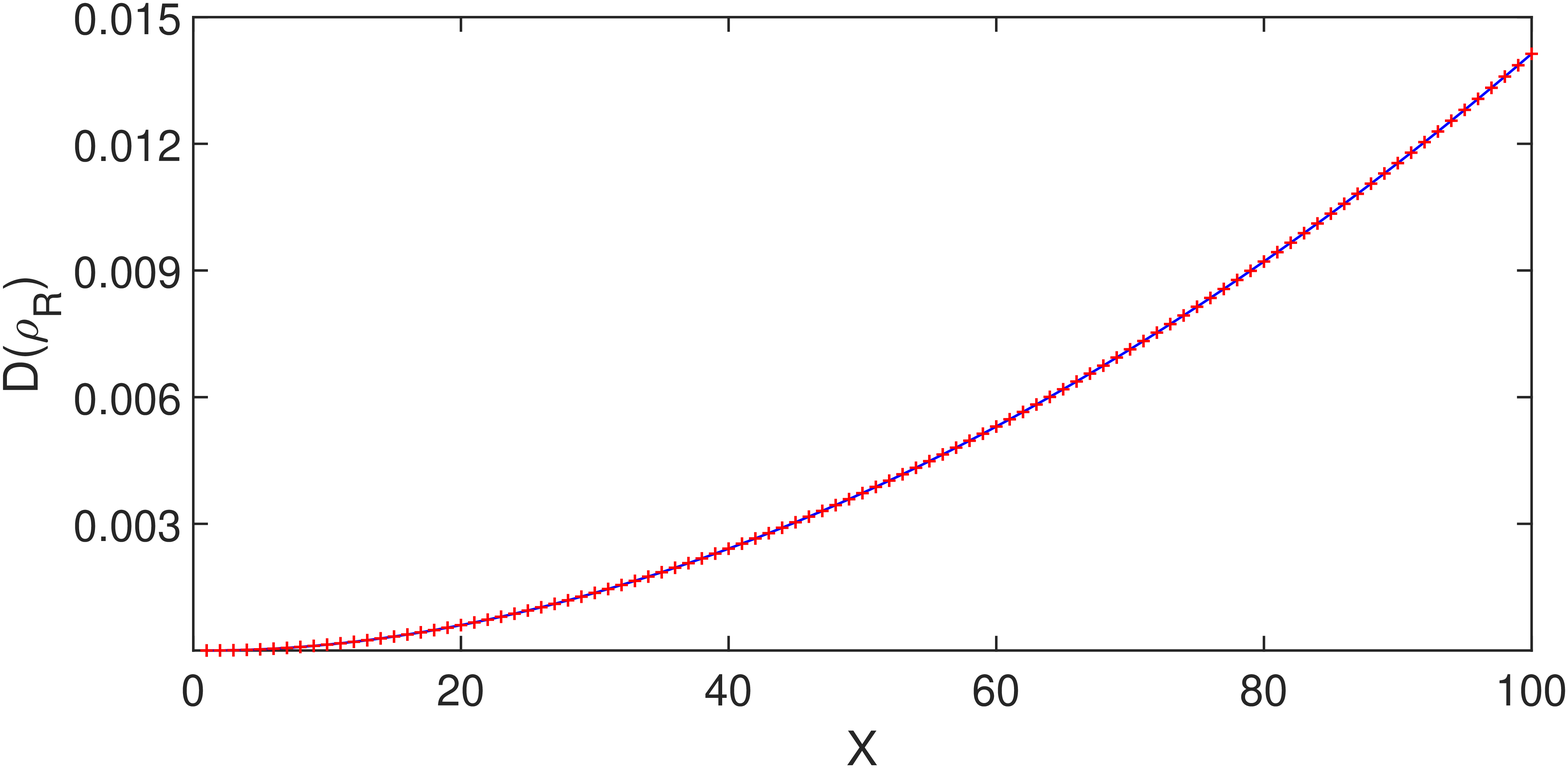}
\caption{(color online) SQD results for "X"-type states $\protect\rho _{R}$
vs. $x$. The solid line corresponds to the strictly analytical expression of
Equation (\protect\ref{dQIUJIE}), while the numerical solutions are marked
with "+" symbols.}
\end{figure}

\textit{Example 4.-The SQD of $(2\otimes 2)$-dimensional general
block-diagonal states}

To demonstrate the validity of our Theorem 3, we consider a state $\rho _{M}$
generated in the same way as that shown in Equation (\ref{G}) but with $G$
replaced by $M$, where the matrix $M$ is constructed as the direct sum of
two randomly generated $2\times 2$ matrices, i.e., $M=M_{1}\oplus M_{2}$,
with%
\begin{equation}
M_{1}=\left(
\begin{array}{cc}
0.3093 & 0.2321+0.0039i \\
0.2321-0.0039i & 0.1885%
\end{array}%
\right) ,
\end{equation}%
and%
\begin{equation}
M_{2}=\left(
\begin{array}{cc}
0.1972 & 0.2075+0.1204i \\
0.2075-0.1204i & 0.3050%
\end{array}%
\right) .
\end{equation}%
We plot the SQD results for such states $\rho _{M}$ as determined both
numerically and analytically. The results are completely consistent with
each other.
\begin{figure}[tbp]
\centering
\includegraphics[width=1\columnwidth,height=1.8in]{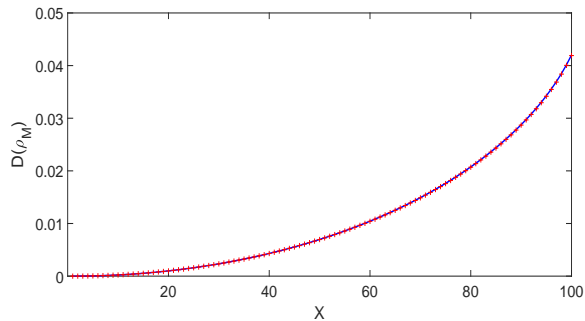}
\caption{(color online) SQD results for block-diagonal states $\protect\rho %
_{M}$ vs. $x$. The solid line corresponds to the strictly analytical
expression of Equation (\protect\ref{discord3}), while the numerical
solutions are marked with "+" symbols.}
\end{figure}

\section{Conclusions and discussion}

We have studied the symmetric quantum discord (SQD) and the symmetric
measurement-induced nonlocality (SMIN) as defined in terms of the quantum
skew information. It is shown that these two quantum correlations do not
change when an additional state is attached to the state of interest. In
particular, we find that the SMIN can be analytically calculated for any
bipartite qubit states and that the SQD can be analytically calculated for
two-qubit states of the "X" type and of the block-diagonal form. Numerical
tests are presented to demonstrate that our analytical expressions are
valid. Since the derivation of analytical expressions for quantifiers of
quantum resources is generally the main challenge in resource theory, we
believe that our results could have broad applications in related studies.

\section*{ACKNOWLEDGMENTS}

This work was supported by the National Natural Science Foundation of China
under Grants No.11775040 and No. 11375036, by the Xinghai Scholar
Cultivation Plan, and by the Fundamental Research Funds for the Central
Universities under Grant No. DUT18LK45.

\appendix

\section{A useful lemma}

\textbf{Lemma: }Let $A$ and $B$ be two nonnormalized qubit density matrices.
There always exists a unitary operation $U$ such that $\left[ UAU^{\dagger }%
\right] _{11}=\left[ UAU^{\dagger }\right] _{22}$ and $\left[ UBU^{\dagger }%
\right] _{11}=\left[ UBU^{\dagger }\right] _{22}$ hold simultaneously.

\textbf{Proof. }Let the unitary matrix $U=\left(
\begin{array}{cc}
\cos \theta & e^{i\varphi }\sin \theta \\
-e^{-i\varphi }\sin \theta & \cos \theta%
\end{array}%
\right) $, and let $A^{\prime }=UAU^{\dagger }$ and $B^{\prime
}=UBU^{\dagger }$. Then, the diagonal entries of $A^{\prime }$ and $%
B^{\prime }$\ read%
\begin{eqnarray}
A_{11}^{\prime } &=&A_{11}\cos ^{2}\theta +A_{22}\sin ^{2}\theta  \notag \\
&&+(A_{12}e^{-i\varphi }+A_{21}e^{i\varphi })\cos \theta \sin \theta ,
\notag \\
A_{22}^{\prime } &=&A_{11}\sin ^{2}\theta +A_{22}\cos ^{2}\theta  \notag \\
&&-(A_{12}e^{-i\varphi }+A_{21}e^{i\varphi })\cos \theta \sin \theta ,
\notag \\
B_{11}^{\prime } &=&B_{11}\cos ^{2}\theta +B_{22}\sin ^{2}\theta  \notag \\
&&+(B_{12}e^{-i\varphi }+B_{21}e^{i\varphi })\cos \theta \sin \theta ,
\notag \\
B_{22}^{\prime } &=&B_{11}\sin ^{2}\theta +B_{22}\cos ^{2}\theta  \notag \\
&&-(B_{12}e^{-i\varphi }+B_{21}e^{i\varphi })\cos \theta \sin \theta ,
\end{eqnarray}%
with $A_{ij}$ and $B_{ij}$, $i,j=1,2$, denoting the entries of the
corresponding matrices. $A_{11}^{\prime }=A_{22}^{\prime }$ and $%
B_{11}^{\prime }=B_{22}^{\prime }$ imply that%
\begin{eqnarray}
&&(A_{11}-A_{22})\cos 2\theta  \notag \\
&=&-(A_{12}e^{-i\varphi }+A_{21}e^{i\varphi })\sin 2\theta ,  \label{q1p1}
\end{eqnarray}%
and%
\begin{eqnarray}
&&(B_{11}-B_{22})\cos 2\theta  \notag \\
&=&-(B_{12}e^{-i\varphi }+B_{21}e^{i\varphi })\sin 2\theta .
\end{eqnarray}%
From these two equations, one can easily find that $p\cos \varphi +q\sin
\varphi =0$, where $%
p=(B_{11}-B_{22})(A_{12}+A_{21})-(A_{11}-A_{22})(B_{12}+B_{21})$ and $%
q=(B_{11}-B_{22})(A_{21}-A_{12})i-(A_{11}-A_{22})(B_{21}-B_{12})i$. Thus, we
have
\begin{equation}
\tan \varphi =-\frac{p}{q}.  \label{pq}
\end{equation}%
By substituting Equation (\ref{pq}) into Equation (\ref{q1p1}), one can
obtain%
\begin{equation}
\tan 2\theta =\frac{A_{11}-A_{22}}{A_{12}e^{-i\varphi
_{2}}+A_{21}e^{i\varphi _{2}}}.
\end{equation}%
Thus, the proof is complete.\hfill $\blacksquare $


\begin{thebibliography}{99}
\bibitem{H1} R. Horodecki, P. Horodecki, M. Horodecki, K. Horodecki, \emph{%
Rev. Mod. Phys.} \textbf{2009}, \emph{81}, 865.

\bibitem{H2} K. Modi, A. Brodutch, H. Cable, T. Paterek, and V. Vedral,
\emph{Rev. Mod. Phys.} \textbf{2012}, \emph{84}, 1655.

\bibitem{H3} A. Streltsov, G. Adesso, M. B. Plenio, \emph{Rev. Mod. Phys.}
\textbf{2017}, \emph{89}, 041003.

\bibitem{B1} C. H. Bennett, H. J. Bernstein, S. Popescu, B. Schumacher,
\emph{Phys. Rev. A} \textbf{1996}, \emph{53}, 2046.

\bibitem{B2} C. H. Bennett, D. P. DiVincenzo, J. A. Smolin, W. K. Wootters,
\emph{Phys. Rev. A} \textbf{1996}, \emph{54}, 3824.

\bibitem{e1} W. K. Wootters, \emph{Phys. Rev. Lett.} \textbf{1998}, \emph{80}%
, 2245.

\bibitem{N2} A. Peres, \emph{Phys. Rev. Lett.} \textbf{1996}, \emph{77},
1413.

\bibitem{Horo} M. Horodecki, P. Horodecki, R. Horodecki, \emph{Phys. Lett. A}
\textbf{1996}, \emph{223}, 1.

\bibitem{N1} G. Vidal and R. F. Werner, \emph{Phys. Rev. A} \textbf{2002},
\emph{65}, 032314 .

\bibitem{ee3} W. D\"{u}r, G. Vidal, and J. I. Cirac, \emph{Phys. Rev. A}
\textbf{2000}, \emph{62}, 062314.

\bibitem{miy} A. Miyake, and F. Verstraete, \emph{Phys. Rev. A} \textbf{2004}%
, \emph{69}, 012101.

\bibitem{CKW} V. Coffman, J. Kundu, and W. K. Wootters, \emph{Phys. Rev. A}
\textbf{2000}, \emph{61}, 052306.

\bibitem{book} G. Alber, T. Beth, M. Horodecki, P. Horodecki, R. Horodecki,
M. R\"{o}tterler, H. Weinfurter, R. Werner, A. Zeilinger, \textit{Quantum
information: An introduction to basic theoretical concepts and experiments},
Springer-Verlag, Berlin Heidelberg, \textbf{2001}.

\bibitem{adp} L. E. Buchholz, T. Moroder, and O. G\"uhne, \emph{Ann. Phys.}
\textbf{2016}, \emph{528}, 278.

\bibitem{G3} S. M. Fei, J Jost, X. Q. Li, G. F. Wang, \emph{Phys. Lett. A}
\textbf{2003}, \emph{310}, 333.

\bibitem{b1} F. Mintert, M. Ku\'{s}, and A. Buchleitner, \emph{Phys. Rev.
Lett.} \textbf{2004}, \emph{92}, 167902.

\bibitem{b2} K. Chen, S. Albeverio, and S. M. Fei, \emph{Phys. Rev. Lett.}
\textbf{2005}, \emph{95}, 040504.

\bibitem{G4} G. Giedke, M. M. Wolf, O. Kr\"{u}ger, R. F. Werner, and J. I.
Cirac, \emph{Phys. Rev. Lett.} \textbf{2003}, \emph{91}, 107901.

\bibitem{ycc} C. S. Yu, and H. S. Song, \emph{Phys. Rev. A} \textbf{2005},
\emph{72}, 022333.

\bibitem{Ost} R. Lohmayer, A. Osterloh, J. Siewert, and A. Uhlmann, \emph{%
Phys. Rev. Lett.} \textbf{2006}, \emph{97}, 260502.

\bibitem{Y3} F. Mintert, \emph{Phys. Rev. A} \textbf{2007}, \emph{75},
052302.

\bibitem{ee2} J. Siewert and C. Eltschka, \emph{Phys. Rev. Lett.} \textbf{%
2012}, \emph{108}, 230502.

\bibitem{ee4} J. Sperling and W. Vogel, \emph{Phys. Rev. Lett.} \textbf{2013}%
, \emph{111}, 110503.

\bibitem{d1} L. Henderson, V. Vedral, \emph{J. Phys. A:Math. Theor.} \textbf{%
2001}, \emph{34}, 6899.

\bibitem{d2} H. Ollivier, W. H. Zurek, \emph{Phys. Rev. Lett.} \textbf{2001}%
, \emph{88}, 017901.

\bibitem{NS} B. Dakic, V. Vedral, and C. Brukner, \emph{Phys. Rev. Lett.}
\textbf{2010}, \emph{105}, 190502.

\bibitem{yzyz} C. S. Yu, and H. Q. Zhao, \emph{Phys. Rev. A} \textbf{2011},
\emph{84}, 062123.

\bibitem{L1} D. Girolami, T. Tufarelli, and G. Adesso, \emph{Phys. Rev. Lett.%
} \textbf{2013}, \emph{110}, 240402.

\bibitem{SY} C. S. Yu, S. X. Wu, X. G. Wang, X. X. Yi and H. S. Song, \emph{%
Europhys. Lett.} \textbf{2014}, \emph{107}, 10007.

\bibitem{RM} M. Rama, \emph{Front. Phys.} \textbf{2016}, \emph{11}, 111404.

\bibitem{OUT} R. Fan, P. Zhang, H. Shen, H. Zhai, \emph{Sci. Bull.} \textbf{%
2017}, \emph{62}, 707.

\bibitem{VON} M. Zhao, T. Ma, T. Zhang, S. M. Fei, \emph{Sci. China-Phys.
Mech. Astron.} \textbf{2016}, \emph{59}, 120313.

\bibitem{QUAN} G. Y. Zhou, L. J. Huang, J. Y. Pan, L. Y. Hu, J. H. Huang,
\emph{Frontiers of Physics} \textbf{2018}, \emph{13}, 130701.

\bibitem{MULT} J. BatleEmail, A. Farouk, O. Tarawneh, S. Abdalla, \emph{%
Frontiers of Physics} \textbf{2018}, \emph{13}, 130305.

\bibitem{EXPE} H. Li, X. Gao, T. Xin, M. H. Yung, G. L. Long, \emph{Sci.
Bull.} \textbf{2017}, \emph{62}, 497.

\bibitem{QUANTU} H. Wang, W. Zheng, N. Yu, K. Li, D. Lu, T. Xin, C. Li, Z.
Ji, D. Kribs, B. Zeng, X. Peng, J. Du, \emph{Sci. China: Phys., Mech. Astron.%
} \textbf{2016}, \emph{59}, 100313.

\bibitem{X1} Y. Huang, \emph{Phys. Rev. A} \textbf{2013}, \emph{88}, 014302.

\bibitem{X2} N. Min, J. Chang, J. Shin, Y. Kwon, \emph{\ Int. J. Theor. Phys.%
} \textbf{2015}, \emph{54}, 3340.

\bibitem{X3} T. Chanda, A. K. Pal, A. Biswas, A. Sen(De), U. Sen, \emph{%
Phys. Rev. A} \textbf{2015}, \emph{91}, 062119.

\bibitem{loca} S. Luo, and S. Fu, \emph{Phys. Rev. Lett.} \textbf{2011},
\emph{106}, 120401.

\bibitem{Wu} S. X. Wu, J. Zhang, C. S. Yu, and H. S. Song, \emph{Phys. Lett.
A} \textbf{2014}, \emph{378}, 344.

\bibitem{iajd} L. Q. Zhang, T. T. Ma, and C. S. Yu, \emph{Phys. Rev. A}
\textbf{2018}, \emph{97}, 032112.

\bibitem{L2} S. Luo, \emph{Phys. Rev. A} \textbf{2008}, \emph{77}, 042303.

\bibitem{G1} S. Luo, and S. Fu, \emph{Phys. Rev. A} \textbf{2010}, \emph{82}%
, 034302.

\bibitem{KM} K. Modi, T. Paterek, W. Son, V. Vedral, and M. Williamson,
\emph{Phys. Rev. Lett.} \textbf{2010}, \emph{104}, 080501.

\bibitem{yb} C. S. Yu, and H. Q. Zhao, \emph{Phys. Rev. A} \textbf{2011},
\emph{84}, 062123.

\bibitem{S1} Q. Chen, C. Zhang, S. Yu, X. Yi, and C. H. Oh, \emph{Phys. Rev.
A} \textbf{2011}, \emph{84}, 042313.

\bibitem{Hu} M. L. Hu, and H. Fan, \emph{New. J. Phys.} \textbf{2015}, \emph{%
17}, 033004.

\bibitem{JOIN} J. Chen, C. Guo, Z. Ji, Y. T. Poon, N. Yu, B. Zeng, J. Zhou,
\emph{Sci. China: Phys., Mech. Astron.} \textbf{2017}, \emph{60}, 020312.

\bibitem{Di1} J. Maziero, L. C. C\'{e}leri, R. M. Serra, \emph{%
arXiv:1004.2082.} \textbf{2010}.

\bibitem{Di2} F. J. Jiang, H. J. L\"{u}, X. H. Yan and M. J. Shi, \emph{%
Chin. Phys. B} \textbf{2013}, \emph{22}, 040303.

\bibitem{Di3} J. Xu, \emph{Phys. Lett. A} \textbf{2012}, \emph{376}, 320.

\bibitem{Y1} C. S. Yu and H. S. Song, \emph{Phys. Rev. A} \textbf{2009},
\emph{80}, 022324.

\bibitem{QQ1} C. S. Yu, Y. Zhang, H. Zhao, \emph{Quantum Inf. Process.}
\textbf{2014}, \emph{13}, 1437.

\bibitem{c1} T. Baumgratz, M. Cramer, and M. B. Plenio, \emph{Phys. Rev.
Lett.} \textbf{2014}, \emph{113}, 140401.

\bibitem{c2} A. Winter and D. Yang, \emph{Phys. Rev. Lett.} \textbf{2016},
\emph{116}, 120404.

\bibitem{c3} E. Chitambar and M. H. Hsieh, \emph{Phys. Rev. Lett.} \textbf{%
2016}, \emph{117}, 020402.

\bibitem{c4} A. Streltsov, S. Rana, P. Boes and J. Eisert, \emph{Phys. Rev.
Lett.} \textbf{2017}, \emph{119}, 140402.

\bibitem{c5} K. B. Dana, M. G. D\'{\i}az, M. Mejatty and A. Winter, \emph{%
Phys. Rev. A} \textbf{2017}, \emph{95}, 062327.

\bibitem{c6} A. Streltsov, S. Rana, M. N. Bera and M. Lewenstein, \emph{%
Phys. Rev. X} \textbf{2017}, \emph{7}, 011024.

\bibitem{c7} C. S. Yu, \emph{Phys. Rev. A} \textbf{2017}, \emph{95}, 042337.

\bibitem{c8} F. G. S. L. Brand\~ao, G. Gour, \emph{Phys. Rev. Lett.} \textbf{%
2015}, \emph{115}, 070503.

\bibitem{c9} S. R. Yang, and C. S. Yu, \emph{Ann. Phys.} \textbf{2018},
\emph{388}, 305.

\bibitem{c10} H. Q. Zhao, and C. S. Yu, \emph{Sci. Reps.} \textbf{2018},
\emph{8}, 299.

\bibitem{QQ3} J. Ma, B. Yadin, D. Girolami, V. Vedral, M. Gu, \emph{Phys.
Rev. Lett.} \textbf{2016}, \emph{116}, 160407.

\bibitem{QQ2} Y. Sun, Y. Mao and S. Luo, \emph{Europhys. Lett.} \textbf{2017}%
, \emph{118}, 60007.

\bibitem{M1} M. Piani, \emph{Phys. Rev. A} \textbf{2012}, \emph{86}, 034101.

\bibitem{zhou} T. Zhou, J. Cui, G. L. Long, \emph{Phys. Rev. A} \textbf{2011}%
, \emph{84}, 062105.

\bibitem{QUANTI} X. F. Qi, T. Gao, F. L. Yan, \emph{Frontiers of Physics}
\textbf{2018}, \emph{13}, 130309.

\bibitem{ACCE} T. Ma, M. J. Zhao, H. J. Zhang, S. M. Fei, G. L. Long, \emph{%
Phys. Rev. A} \textbf{2017}, \emph{95}, 042328.

\bibitem{skew} E. P. Wigner and M. M. Yanase, \emph{Proc. Natl. Acad. Sci.
U. S. A.} \textbf{1963}, \emph{49}, 910.

\bibitem{skew1} E. H. Lieb, \emph{Adv. Math.} \textbf{1973}, \emph{11}, 267.

\bibitem{skew2} S. Luo, \emph{Proc. Am. Math. Soc.} \textbf{2004}, \emph{132}%
, 885.

\bibitem{skew3} Z. H. Ma, Z. H. Chen, S. M. Fei, \emph{Sci. China-Phys.
Mech. Astron.} \textbf{2017}, \emph{60}, 010321

\bibitem{ruili} G. H. Golub, C. F. Van Loan, \textit{Matrix Computations.
2nd ed.} The John Hopkins University Press, Baltimore, \textbf{1989}.
\end{thebibliography}
\end{document}